\documentclass[10pt,journal,compsoc]{IEEEtran}
\usepackage[utf8]{inputenc}
\usepackage{tabularx}
\usepackage{graphicx,color}
\usepackage{balance}
\usepackage{array}
\usepackage{varwidth}
\usepackage{pgfplots}
\usepackage{comment}
\usepackage[normalem]{ulem}

\newcolumntype{P}[1]{>{\centering\arraybackslash}p{#1}}
\begin{document}

\title{Integrating Digital Twin and Advanced Intelligent Technologies to Realize the Metaverse}

\author{
\IEEEauthorblockN{
\textbf{Moayad Aloqaily}, \textit{Senior Member, IEEE}, 
\textbf{Ouns Bouachir}, \textit{Member, IEEE},
\textbf{Fakhri Karray}, \textit{Fellow, IEEE},
\textbf{Ismaeel Al Ridhawi},    \textit{Senior Member, IEEE},
\textbf{Abdulmotaleb El Saddik},        \textit{Fellow, IEEE}
} 
\IEEEcompsocitemizethanks{

\IEEEcompsocthanksitem M. Aloqaily, F. Karray and A. El Saddik are with the Machine Learning Department, Mohamed Bin Zayed University of Artificial Intelligence (MBZUAI), UAE. \protect E-mails: \{moayad.aloqaily;Fakhry.karray;Abdulmotaleb.ElSaddik\}@mbzuai.ac.ae% <-this % stops an unwanted space

\IEEEcompsocthanksitem O. Bouachir is with Zayed University, UAE. \protect E-mail: ouns.bouachir@zu.ac.ae

\IEEEcompsocthanksitem I. Al Ridhawi is with Kuwait College of Science and Technology, Kuwait. \protect E-mail: i.alridhawi@kcst.edu.kw}}

%author{
 %       ~\textbf{Moayad Aloqaily},~\IEEEmembership{Senior Member,~IEEE,}
  %      ~\textbf{Ouns Bouachir},~\IEEEmembership{Member,~IEEE,}
   %     ~\textbf{Fakhri Karray},~\IEEEmembership{Fellow,~IEEE,}
    %    ~\textbf{Ismaeel Al Ridhawi},~\IEEEmembership{Senior Member,~IEEE}
     %   ~\textbf{Abdulmotaleb El Saddik}% <-this % stops a space
%}
%\author{Moayad Aloqaily}
%\affil{Mohamed Bin Zayed University of Artificial Intelligence (MBZUAI), UAE}

%\author{Ouns Bouachir}
%\affil{Zayed University, UAE}

%\author{Fakhri Karray}
%\affil{Mohamed Bin Zayed University of Artificial Intelligence (MBZUAI), UAE}

%\author{Ismaeel Al Ridhawi}
%\affil{Kuwait College of Science and Technology (KCST), Kuwait}

%\author{Abdulmotaleb El Saddik}
%\affil{University of Ottawa, Canada and Mohamed Bin Zayed University of Artificial Intelligence (MBZUAI), UAE}

%\IEEEcompsocitemizethanks{\IEEEcompsocthanksitem 

%\par \noindent M. Aloqaily, F. Karray, A. El Saddik are with MBZUAI, UAE. \protect E-mails: maloqaily@ieee.org% <-this % stops an unwanted space

%\par \noindent I. Al Ridhawi is with University of Ottawa, Canada. \protect E-mail: ismaeel.alridhawi@uottawa.ca

%\par \noindent O. Bouachir is with ZU, UAE. \protect E-mail: ouns.bouachir@zu.ac.ae% <-this % stops an unwanted space
%}}

%\end{comment}
%\markboth{IEEE Network,~Vol.~x, No.~x}%
\maketitle

\begin{abstract}
The advances in Artificial Intelligence (AI) have led to technological advancements in a plethora of domains. Healthcare, education, and  smart city services are now enriched with AI capabilities. These technological advancements would not have been realized without the assistance of fast, secure, and fault-tolerant communication media. Traditional processing, communication and storage technologies cannot maintain high levels of scalability %and end-to-end service qualities 
and user experience for immersive services. The metaverse %which has recently gained popularity among researchers, 
is an immersive three-dimensional (3D) virtual world that integrates fantasy and reality into a virtual environment using advanced virtual reality (VR) and augmented reality (AR) devices. Such an environment is still being developed and requires extensive research in order for it to be realized to its highest attainable levels. In this article, we discuss some of the key issues required in order to attain realization of metaverse services. We propose a framework that integrates digital twin (DT) with other advanced technologies such as the sixth generation (6G) communication network, blockchain, and AI, to maintain continuous end-to-end metaverse services. This article also outlines requirements for an integrated, DT-enabled metaverse framework and provides a look ahead into the evolving topic.

\end{abstract}

    
%\IEEEpeerreviewmaketitle
\enlargethispage{10pt}

%\chapterinitial{The introduction}
\section{Introduction}
The term "metaverse" is being heralded by researchers and analysits as the next evolutionary technology. Given that the term is still immature, there is no clear consensus as to what this technology will evolve to. Currently, the term is being used to refer to three dimensional (3D) virtual environments that features a hybrid world of real and imitational objects such as avatars and other forms of digital objects. It enables the physical world to exist within a virtual world. For example, transactions conducted in the real world also exist in the virtual world. The metaverse encompasses the physical space into cyberspace, where people can interact, socialize, construct objects, and engage in transactions with one another using avatars. Transaction purchases can be exchanged using crypto-currencies. It is thus an evolved Internet that is supported by advanced technologies such as augmented reality (AR), virtual reality (VR), mixed reality (MR), and other forms of extended reality (XR) \cite{chapter}. Although the metaverse is now being investigated in both academia and industry, fundamental challenges and obstacles that need to be considered in order to realize the metaverse remains to be explored.

The three core requirements needed to realize the metaverse are the \textit{i)} cloning of tangible and intangible objects into the virtual environment, \textit{ii)} self-configuring and managing the virtual environment, and \textit{iii)} maintaining trust and authenticity between virtual objects and transactions. The integration and attainment of those three requirements will require significant amount of data collection, gathering, analysis, cloning, and transmission. Although there have been a number of attempts to prototype metaverse applications and services \cite{intro1}, with the existing communication and networking infrastructure, it becomes difficult to realize the metaverse at full-scale. With the advancements in next-generation networks (NGN), especially the sixth-generation (6G) communication network, data transmission, communication, and networking can significantly enhance the experience quality for immersive metaverse services and applications \cite{intro2}. Indeed, in such a system, the network has an important role to guarantee the high transmission capacity and accuracy with the least delay for the  massive amounts of heterogeneous data. This must be achieved while ensuring the synchronization and interaction of the metaverse experience. As the enabler of these advanced technologies, 6G is a key pillar for metaverse services \cite{conclusion1}.

Object cloning from the physical world into the virtual world necessitates that objects be accurately resembled. Digital twin (DT) is a paradigm that has emerged lately to enable the representation of a digital mirror for a physical entity \cite{intro3}. The twins evolve synchronously throughout the lifetime of the physical entity, and can evolve separately in the metaverse, thanks to the real-time data gathering and the continuous advanced processing. For instance, during the lifetime of a person in the real-world, both the real person and the digital twin are synchronized both in the real and virtual worlds. But, after the person's lifetime, its digital twin can continue to exist and evolve with the support of advanced Artificial Intelligence (AI) solutions. In essence, the realization of the metaverse is highly correlated with the advancements in DT. Not only should metaverse avatars dimensionally resemble their physical counter-parts (using advanced AI-supported image capturing devices and algorithms), but it should also have abstract resemblance in terms of behaviour, actions, and decision making. This can only be achieved using AI-supported DT techniques that clone the data and perform data analysis intelligently.

The involvement of AI-enabled solutions in the form of machine learning (ML), deep learning (DL), and federated learning (FL), is significantly important in order to maintain a high-level of self-management of the metaverse. Given that transactions and interactions made in the metaverse are captured through XR devices, movement and characteristics of both real and virtual world objects must have high levels of precision. This is only achievable with AI-enabled algorithms. Moreover, self-configuration extends to real-world objects and actions. Network and device management requires enhanced AI-supported solutions to maintain the continuous availability of end-to-end metaverse applications and services.

Given that the metaverse incorporates large volumes of interactions and transactions between participants, trust and integrity issues are of a main concern. Maintaining a centralized solution for trust and transaction management is inefficient and is not adaptable at the current state. It is envisioned that most of the transactions conducted in the metaverse will be in the form of crypto-currencies \cite{intro4}. Thus, a decentralized solution for transaction authentication is necessary. Blockchain is a promising technology that can be integrated into the metaverse to maintain decentralization, transparency, and immutability \cite{intro5}. It is expected that blockchain will enable and enforce trust and accountability in this virtual ecosystem. Not only is the authenticity of transactions secured through blockchain, but also the integrity, privacy, and reputation of participants are also maintained. 

Some recent work focused on the performance of DT in the metaverse and proposed solutions to enhance the DT security and quality of experience \cite{ relatedWork1, relatedWork2, relatedWork3}. Such solutions focus on DT service performance as a single entity in the metaverse. However, these solutions consider that DT is one of many pillars of the metaverse and discuss the role of DT and its interactions with the other metaverse enablers.

With that said, this article aims at introducing a conceptual framework of a DT-enabled metaverse system. The solution aims at integrating a number of intelligent solutions such as DT, blockchain, AI, and XR to provide an enhanced metaverse experience while maintaining high levels of service quality and efficient resource management. This article also discusses some of the requirements needed to realize the framework and paves the road ahead for interested researchers.

% About 6G/DT - the history of integrating
% Motivation - Virtualization in 6G
% Taxonomy of the survey.
% Comparison with another similar work – Table
% How is this magazine different from others – the main delta?
% Contribution of the magazine

%\section{Integration of the DT with Futuristic Technologies}
% Introduce it
% Advantages and challenges
% Minimum 1 figure and focus more on tables to summarize the contents

%\section{Requirements}
%Requirement for the integration to happen?

%\section{Existing Architecture}
%Implementation / tools / technologies available
%\vspace{-3mm}
\section{A Conceptual Framework for the Metaverse}
%\subsection{DT-6G for Metaverse}
Being one of the most exciting futuristic services, the metaverse has an undeniable potential to change technology on a global scale and push forward the advancement of cities, industries, and services. The metaverse can be defined as the gathering of data and information surrounding users and objects, collected from various locations and devices (\textit{e.g.,} homes, work, and IoT devices), and presented in an immersive manner while being accessible through XR, including AR, VR, and mixed reality (MR), for a more tangible, visual, and interactive experience. In other words, the metaverse offers a new way to interact with the processed and collected data.
The metaverse is one of the services that can gain several advantages from DT. Indeed, the creation of DT for metaverse services would enable continuous monitoring of end-user requests and provide the needed resources to enhance services and achieve adequate levels for quality of experience (QoE). Such services would take into consideration the various devices, concepts, and operations that are included in this process, most importantly network resources. On the contrary, DT for real objects can be used to form a real-word metaverse for a realistic immersive experience.
%\vspace{-2mm}
\subsection{DT for Metaverse Services}
To provide the expected immersive services, metaverse relies on various operations, namely, data collection, transmission, manipulation, and generation. The metaverse also relies on a plethora of devices such as participants’ mobile phones, cameras, helmets, and edge nodes. More importantly, the metaverse relies on a diverse set of technological advancements to build a real-world metaverse, including, next-generation networking and communications, ML, AI, and DT. DT is a concept that provides several advantages for monitoring entities and operations of such complex ecosystems. To create the DT for the metaverse service, all DTs of all the involved operations, concepts, and entities should work together to monitor and provide a comprehensive and accurate analysis of the provided service. For instance, the DT of the metaverse service should communicate with the wireless network DT (\textit{i.e.,} which includes the DT for the various network operations and infrastructure), given that the network performance has an influential impact on the quality of the delivered service. On the other hand, the projection of the current and predicted metaverse participants’ requests on the DT helps in adapting for the network resource allocation to reach the highest levels of QoS and autonomy.
%\vspace{-2mm}
\subsection{DT for Real-World Metaverse}
The metaverse can provide a new dimension towards the digital transformation of smart cities. It can enhance the current services and provide a new range of applications. The creation of the real-world metaverse has a big impact on the advancement of services, products, cities, and countries, as it enables the gathering of the utility, the context, and the relevance of both real and virtual life. In other words, far from entertainment services, a metaverse experience that duplicates the real world can provide advanced analysis of various situations that can occur in real life and enables in understanding the footprints of any system and making accurate daily choices. 

In this context, the metaverse relies on DT to visualize and construct the real physical aspects of the objects forming the virtual environment. Using the real-time collected data and extracted information of the DT, the digital clone of the real world can be created in the metaverse, reproducing the same real features of the objects, and providing realistic responses based on the physical characters collected and analyzed by the DT. As such, the visualization of the DT outcome is more significant and comprehensive in such an environment. Building a real-world metaverse provides a new advanced dimension to visualize and interact with information and allows for monitoring systems in various sectors such as healthcare, construction, shopping, and tourism. Additionally, DT in the metaverse can provide several benefits to forensics and safety for smart cities.

\begin{figure*}[ht!]
    \centering
    \includegraphics[width=\textwidth]{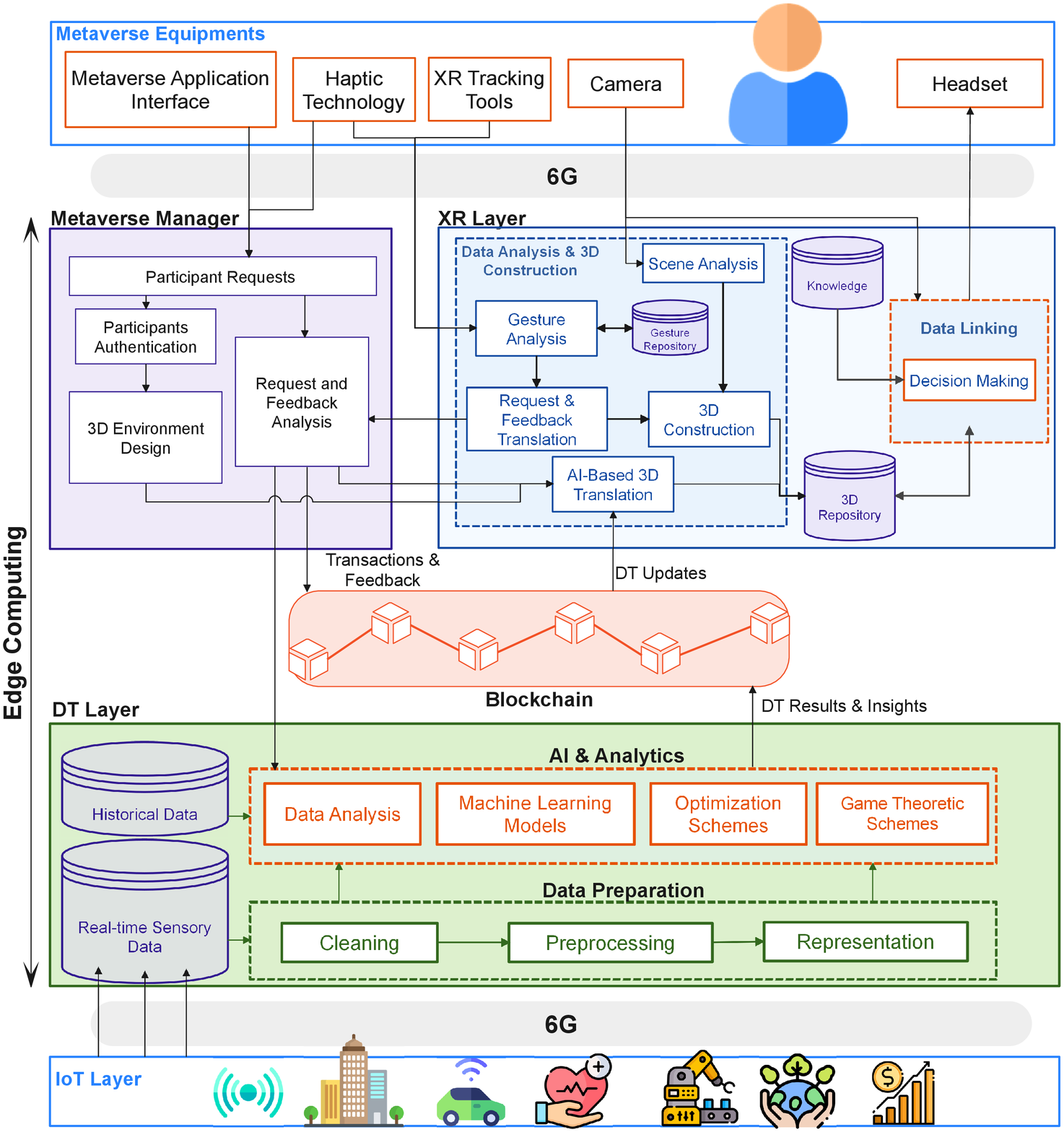}
    \caption{A detailed overview of the enhanced DT-enabled metaverse framework. Immersive XR services are realized through cooperative entities over the edge. Data exchange and communication is maintained over 6G communication.}
    \label{fig:platform}
    %\vspace{-5mm}
\end{figure*}

\subsection{Metaverse Requirements}
To build the real-world metaverse, the DT of every single entity, concept, or operation should be involved in the created 3D environment. Thus, a huge amount of various types of data are highly needed in such a system, including Internet of Things (IoT) data and high-resolution videos (\textit{e.g.,} 8k videos). Significant amounts of data (\textit{i.e.,} terabytes) need to be exchanged to realize real-word metaverse. This data should go through AI and ML-based processing methods to build the DT and visualize its 3D representation. In addition, the metaverse provides a flexible and natural way of interacting with the created objects by allowing the participants to use their real-life gestures such as grabbing, moving, and zooming. 

To this end, dedicated tools and approaches to track user gestures should be used, thus extracting other types of data that have to be analyzed and translated into requests and feedback. As a result, this helps to achieve the highest level of interaction with the metaverse objects. In other words, various tasks should be performed starting from data collection, transmission, analysis, processing, and finally sharing the output with all the metaverse participants. All these tasks have to be performed with the highest precision and the least delay to achieve total synchronization among the various participants and to provide immediate responses from the 3D objects as in real life. 

The real-world metaverse is based on the cooperation and communication between diverse DTs for each object in the 3D environment. The data accessed and exchanged in this regard creates several challenges related to data accessibility, availability, and integrity. Moreover, data can have a significant impact on the system performance in terms of delay, responsiveness, and accuracy. To provide the needed requirements, diverse entities from several stakeholders should cooperate to reach the expected QoS. Such cooperation represents one of the biggest challenges in terms of the complexity of the methods to be used, algorithm standardization, transparency, and trust.

\section{Enhanced DT-enabled Metaverse Framework}
%\subsection{Framework Architecture}
%The real-world metaverse is based on the cooperation and communication between the DTs of each object to recreate their real physical features in the 3D environment. The data accessed and exchanged in this regard creates several challenges related to its accessibility, availability, interoperability and integrity as it has a big impact on the system delay,  responsiveness, accuracy, and overall performance.
%On the other hand, to provide the needed requirements, diverse entities from several stakeholders should cooperate to reach the expected QoS. This feature represents one of the biggest challenges in such a situation related to the methods and algorithms standardization,  transparency, and trust.
    For such a complex ecosystem, the designed platform has numerous challenges in order to meet the aforementioned requirements and ensure the interoperability between its various components in a fast and secure manner. The various modules of the proposed DT-enabled platform %, depicted in Figure \ref{fig:fig1} 
    are integrated into a distributed architecture, where various edge nodes cooperate % with the cloud 
    to provide the highest levels of QoE and QoS. This cooperation is maintained by different system components and communicated over an ultra-fast infrastructure, namely, 6G. Both DT and XR modules will run over the edge nodes cooperatively. The management of those modules must also be maintained at the edge. Data provision and collection from the real-world must consider 6G communication to attain high service and experience levels in metaverse applications.

%\begin{figure}[t]
  %  \centering
 %   \includegraphics[width=0.49\textwidth]{Figures/fig1.eps}
 %   \caption{High-level overview of a DT-enabled metaverse system. Immersive XR services are realized through cooperative entities over the cloud and edge. Data exchange and communication is maintained over 6G communication.}
%    \label{fig:fig1}
%\end{figure}

Figure \ref{fig:platform} provides a detailed overview of the proposed DT-6G metaverse platform. The proposed platform is composed of various layers and components, namely, \textit{i)} IoT layer, \textit{ii)} DT layer, \textit{iii)} XR layer, \textit{iv)} metaverse manager, and \textit{v)} blockchain. Each layer focuses on a set of given tasks. The communication between the various modules of these layers and components is based on the 6G network to provide the required QoS for the exchanged data.

\subsection{IoT Layer}
The IoT layer provides the needed infrastructure to allow for data gathering and to provide results in the form of actions. As such, it includes sensors, devices, methods, and  approaches to collect and manage any type of data that can be useful in building the DT for an object, service, or system in the real world. 

\subsection{Digital Twin Layer}
The DT layer encloses the various algorithms and mechanisms used to build the DT of any object. It uses AI and ML-based approaches with optimization and game theory schemes to analyze the gathered data and compare results to previous conclusions and insights. As a result, the AI and ML-based approaches help adjust and build a more comprehensive pattern of the focal system of the DT. The resulting patterns are very useful in understanding the real-time performance of the system, simulating planned modifications, and predicting possible events. The DT layer stores the digital information of any created digital replication in the blockchain to ensure the fastest accessibility to the data, mainly, data related to the various created DT. Moreover, the stored digital information helps to obtain high levels of trust and accuracy between the real object and its DT.

\subsection{Metaverse Manager}
This module represents the interface between the DT layer and the XR layer, as well as between the participants and the metaverse. Its main task is the design of the metaverse environment based on the participant request. The needed information and responses are retrieved from the DT layer and communicated through the blockchain to the XR layer, where the 3D construction is performed. The DT models are used to build the metaverse 3D objects and operations in a realistic manner by reproducing their real physical characteristics. With AI, these 3D duplications enable real response and interaction with the metaverse participants, as they would in real-life. The metaverse manager provides a dedicated application programming interface (API) used to simplify the interaction between the XR and DT layers, while providing a highly interactive immersive experience for the participants. Additionally, the metaverse manager handles all the participant requests and transactions during the immersive experience and stores all the related information into the blockchain to achieve the highest level of transparency, accounting and, interoperability. 
  
\subsection{XR Layer}
This layer contains all the operation and algorithms needed to build a 3D environment of the metaverse, based on the information received from the DT layer. This layer also provides a flexible and natural interactive experience with the 3D objects. Using dedicated tools based on haptic technology and dedicated sensors (\textit{e.g.,} sensors to detect eye movement) to track user gesture (\textit{e.g.,} grabbing, moving, and zooming) the XR modules analyze the detected motions and translate them into requests to be performed on the virtual objects. The realistic object responses are obtained by the metaverse manager from the DT information stored in the blockchain by the DT layer.
   
%\subsection{Integrating Blockchain with the DT-enabled Metaverse Framework}
\subsection{Blockchain}
To build the real-world metaverse, a huge amount of information should be extracted and combined to build the digital clone of real-world objects using DT. Thus, the accessibility of the data and its accuracy represent an important factor in order to provide the highest level of QoE. The blockchain adopted into the framework provides several advantages for the DT-metaverse. It provides a mechanism to prove the identity of each DT in regard to their real object, and thus, guarantees the accuracy of the collected data and processed information. In addition, the distributed feature of the blockchain allows the metaverse applications to globally access the needed data without a third-party for faster and more transparent data accessibility. Ultimately, it simplifies the creation of the 3D environment based on several DTs and enables various participants from different locations to join the same virtual world. Also, the various transactions and metaverse service related information are added to the blockchain and shared with the DT layer to adjust the created digital models in transparency.

\begin{figure}
       \centering
       \includegraphics[width=0.49\textwidth]{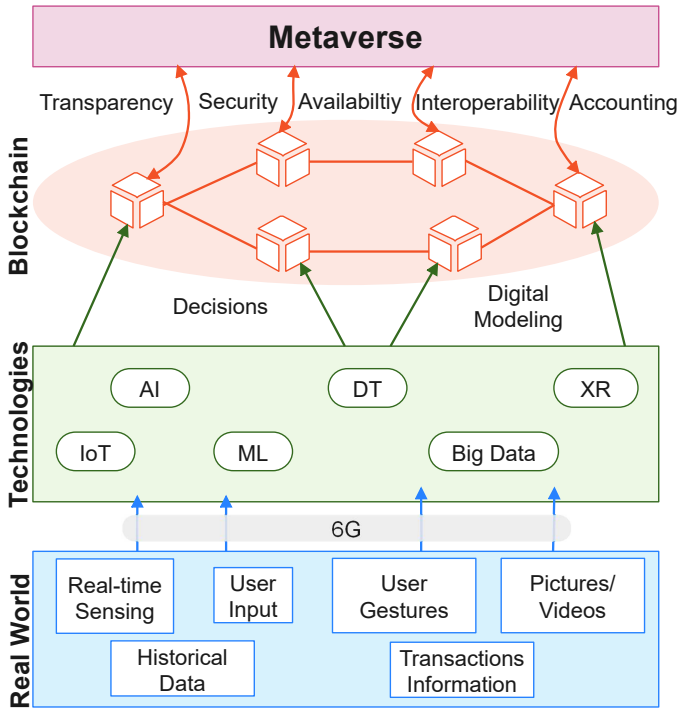}
       \caption{Blockchain in the enhanced DT-enabled metaverse framework.}
       \label{fig:BCl}
       %\vspace{-5mm}
   \end{figure}

Figure \ref{fig:BCl} depicts the data stored into the blockchain in the form of decisions, models, responses and the context of the immersive environment. This information must be authenticated to ensure the integrity of the shared data. The figure shows that the various technologies are capable of interacting with the blockchain to enhance data interoperability. To generate the required metaverse experience, the needed information is extracted from the blockchain close to the participants where the metaverse service is requested and consumed. The blockchain provides several advantages in this context, including data transparency, security, availability, accountability and interoperability.

%\vspace{-2mm}
\section{Real-world Metaverse Framework Operation}
The real-world metaverse is realized through the cooperation of various technologies including DT, XR, 6G, and the blockchain. To provide a highly interactive immersive experience in the metaverse, the proposed framework requires continuous interaction between the different layers and components. This interaction can be divided into several steps:
%\vspace{-2mm}
\begin{itemize}

  \item The first step is to create a DT of the various objects, operations, and services. Different types of data should be collected from the real world and forwarded in real-time through the 6G network to the designated edge nodes. The selection of the edge nodes should be performed based on an adaptive task offloading scheme that takes into consideration the capability, the load, and the location of the participants \cite{GlobeCom21}. The collected data, at each edge node, should then go through a pre-processing phase for extracting and cleaning  the useful data that can be used to build the DT. Then, the AI and ML-based models of the DT layer process and analyze the data to build/update the DT and adjust its features. The obtained information is then added to the blockchain after its validation by the various miners (\textit{i.e.,} the edge nodes).

\item Once the participants initiate a request to join the metaverse, the request is received by the metaverse manager layer, where it is processed to extract the requested features for designing the 3D environment, and connecting with the other participants.
\item The creation of the 3D environment is the task of the XR layer, which extracts the DT information of the designed environment from the blockchain and carries out the needed tasks for data linking and 3D construction. A data tracking bridge is then created between the blockchain, the XR, and the DT layers to forward the DT information and update the metaverse environment in real-time. The obtained visual 3D elements are then forwarded to the various participants through the 6G network.
\item The XR layer allows the participants to interact in a flexible and natural manner using dedicated tracking tools such as the handheld trackers or the sensors embedded in the XR headset. The gestures of the participants are captured, analyzed, and translated into requests that are then forwarded to the metaverse manager layer where they are processed. Using the AI-based approaches developed in the DT layer, realistic responses of the objects are generated and injected into the metaverse through the XR layer.
\end{itemize}
%\vspace{-3mm}
%How will we design and implement it such system
%What are the requirement of metaverse and how DT-6G is fundamental in this environment 
\section{Concluding Remarks and the Way Ahead}
Given that the concept of the metaverse is evolving with technology, it is envisioned that the metaverse will become more than just an ecosystem for avatars and virtual interactions. The metaverse will become a cyberspace for content creation, virtual economy, social interactions, and service provisioning that will have an effect on the real world and objects. Imagine conducting a virtual purchase transaction in the metaverse that would affect the behaviour of physical transactions in the real-world. Similarly, virtual social interactions in the metaverse might result in a behaviour change of people in the real world. In essence, it is expected that the metaverse ecosystem will evolve in the following areas:
%\vspace{-2mm}
\begin{itemize}
    \item \textbf{Personification}: appearance, design, user perception, avatar interaction, and avatar behaviour.
    \item \textbf{Content}: creation, trading, authoring, ownership, censorship, culture, and collaboration.
    \item \textbf{Economy}: transactions, trading, commerce, governance, property ownership, currency establishment, and exchange.
    \item \textbf{Society}: a whole new social experience will be introduced, and thus, there needs to be guidelines and policies that are set in terms of diversity, equality, threats, and cyber-bullying.
    \item \textbf{Trust and Security}: transaction authenticity, resource ownership, biometric information, avatar authenticity, data integrity, and auditability.
\end{itemize}
 
 The provisioning of the metaverse cannot be achieved without the integration of numerous technologies. As discussed in this article, AI, DT, XR, 6G, and blockchain are key elements to  providing an enhanced experience in the metaverse. There still remains significant work to be done in the following areas to realize end-to-end metaverse services:%\vspace{-2mm}
 \begin{itemize}
     \item \textbf{Communication Network}: traditional communication and networking infrastructures will not be able to cope with excessive data communication. In essence, NGNs, including 6G, must consider the metaverse requirements in terms of QoS, QoE, congestion control, network slicing, and network virtualization \cite{conclusion1}.
     \item \textbf{Edge Computing and Storage}: cloud solutions will require significant support from distributed edge computing paradigms for data storage, analysis, federated learning, and real-time service provisioning.
     \item \textbf{Artificial Intelligence}: All elements involved in the provisioning of metaverse services and applications will require the integration of AI. XR devices need to incorporate lightweight AI solutions to support advanced projection and hologram creation mechanisms. Plug-and-play AI, or PnP-AI, will play a significant role in the adaptation of ML into end-devices \cite{pnpAI}. A significant portion of the processing will be performed on PnP-AI-supported devices to maintain efficient levels of data communication on the network. Moreover, AI needs to be incorporated into other domains such as networking, blockchain, IoT devices, sensors, and edge devices.
     \item \textbf{Blockchain}: decentralized data, transaction authentication, and storage solutions, such as blockchain, require a leap towards efficient scalability. The current form of blockchain still has some drawbacks including complexity and total anonymity. Blockchain should be incorporated into more distributed solutions to provide more scalability. Hierarchical blockchain is one example of a blockchain structure that distributes the control and management of data not only among different entities, but also integrates edge and cloud devices to participate in the consensus process \cite{blockchain}.
     \item \textbf{Extended Reality Devices}: XR devices in their current form are considered early-stage equipment for the envisioned metaverse. Sensors and virtual reality space convergence devices must provide advanced, simple, and compact solutions for the provisioning of holographic images, signal processing, recognition, and real-time rendering.
     \item \textbf{Management Technology}: to ensure enhanced experience and service of metaverse users, resource, session, and energy management are extremely important. Advanced and intelligent resource discovery, addressing, and allocation solutions are areas that must be improved. Single and multi-session management are other areas that must be considered. Lastly, and most importantly, nowadays, energy usage and efficiency are of great concern. No doubt that metaverse services and applications will require increased energy consumption. With that said, it is important to adopt green and energy-efficient equipment, networking and service delivery solutions \cite{green}.
 \end{itemize}
 %\vspace{-2mm}

The evolution of the metaverse will without a doubt come to existence as it has with the evolution of other technologies. Visions thought about decades ago are now a reality. Similarly, the visions that we layout today will become true sooner or later. Researchers from both academia and industry will need to enhance current solutions, especially in the area of AI, so that one day the integration between both reality and the metaverse will become joint.
For the future work, we plan to implement the proposed platform and study its performance and impact on metaverse services. This implementation will help researchers in studying new related approaches and concepts. 
%\vspace{-3mm}
\balance
\bibliographystyle{IEEEbib}
\bibliography{general}

\end{document}